*Article*

# *In Situ* Electrochemical SFG/DFG Study of CN⁻ and Nitrile Adsorption at Au from 1-Butyl-1-methyl-pyrrolidinium Bis(trifluoromethylsulfonyl) Amide Ionic Liquid ([BMP][TFSA]) Containing 4-{2-[1-(2-Cyanoethyl)-1,2,3,4-tetrahydroquinolin-6-yl]diazenyl} Benzonitrile (CTDB) and K[Au(CN)$_2$]

**Benedetto Bozzini [1],\*, Bertrand Busson [2], Audrey Gayral [2], Christophe Humbert [2], Claudio Mele [1], Catherine Six [2] and Abderrahmane Tadjeddine [1]**

[1] Department of Innovation Engineering, University of Salento, via Monteroni, 73100 Lecce, Italy
[2] Physical Chemistry Laboratory, Paris-Sud University, CNRS, Bat 201P2, 91405 Orsay, France

\* Author to whom correspondence should be addressed; E-Mail: benedetto.bozzini@unisalento.it.

**Abstract:** In this paper we report an *in situ* electrochemical Sum-/Difference Frequency Generation (SFG/DFG) spectroscopy investigation of the adsorption of nitrile and CN⁻ from the ionic liquid 1-butyl-1-methyl-pyrrolidinium bis(trifluoromethylsulfonyl) amide ([BMP][TFSA]) containing 4-{2-[1-(2-cyanoethyl)-1,2,3,4-tetrahydroquinolin-6-yl]-diazenyl}benzonitrile (CTDB) at Au electrodes in the absence and in the presence of the Au-electrodeposition process from K[Au(CN)$_2$]. The adsorption of nitrile and its coadsorption with CN⁻ resulting either from the cathodic decomposition of the dye or from ligand release from the Au(I) cyanocomplex yield potential-dependent single or double SFG bands in the range 2,125–2,140 cm$^{-1}$, exhibiting Stark tuning values of ca. 3 and 1 cm$^{-1}$ V$^{-1}$ in the absence and presence of electrodeposition, respectively. The low Stark tuning found during electrodeposition correlates with the cathodic inhibiting effect of CTDB, giving rise to its levelling properties. The essential insensitivity of the other DFG parameters to the electrodeposition process is due to the growth of smooth Au.

## 1. Introduction

Electrochemistry in room-temperature ionic liquids (RTIL) is a rapidly developing topic, with prospective applications in electrodeposition and energetics [Li-ion batteries, supercapacitors and proton-exchange membrane fuel-cells (PEMFC)] [1]. Despite the abundance of recent literature, spectroelectrochemical methods are seldom used, at the time of this writing, the following approaches have been described: Fourier-transform infrared (FT-IR) spectroscopy [2], surface-enhanced infrared absorption (SEIRA) spectroscopy [3], surface-enhanced Raman scattering (SERS) spectroscopy [4] and sum-frequency generation (SFG) spectroscopy [5–8]. Spectroelectrochemistry during metal plating provides useful information on the growth interface and a range of approaches has been proposed to achieve information on the chemical composition, electronic structure and adsorption at the dynamic electrochemical interface. In particular, SFG has proved particularly informative because it combines utmost surface sensitivity (bulk signal is not allowed within the electric dipole approximation) and single state capability (steady-state electrochemical conditions are able to yield a high signal-to-noise ratio, at variance e.g., with FT-IR and SERS) with sensitivity to both vibrational and electronic structure of the interface [8,9]). A particular advantage of *in situ* spectroelectrochemistry during electrodeposition processes is the possibility of monitoring the state of additives at the growing interface, yielding molecular-level information that can be directly correlated to phenomenological or ex situ quality indicators of the performance of agents, such as brighteners and levellers, that are of paramount importance in industrial plating processes. Among plating additives, 4-{2-[1-(2-cyanoethyl)-1,2,3,4-tetrahydroquinolin-6-yl]diazenyl} benzonitrile (CTDB) has been recently proved to be highly diagnostic of a quite comprehensive range of interfacial processes in which levellers are involved [10,11].

In this paper we propose an investigation of the potential-dependent electrodic behaviour of CTDB added to 1-butyl-1-methyl-pyrrolidinium bis(trifluoromethylsulfonyl) amide ([BMP][TFSA]) electrolytes in contact with Au electrodes, comparing the interfacial action of this model leveller in the absence and in the presence of ongoing Au electrodeposition. Cathodic operation of CTDB at sufficiently negative polarisations and reduction of K[Au(CN)$_2$] independently lead to the release of CN$^-$ in the electrolyte. Since this pseudohalide tends to adsorb strongly on Au, coadsorption of CTDB with CN$^-$ takes place, resulting in a rich interfacial compositional scenario.

## 2. Results and Discussion

*2.1. Cyclic Voltammetry*

Cyclic voltammograms (CV) of Au in contact with [BMP][TFSA]-based electrolytes, without and with CTDB and K[Au(CN)$_2$] are shown in Figure 1. The CV measured with pure RTIL is essentially the same as that reported in [7,8]. Anodic and cathodic decomposition reactions of RTIL occur at about 2.0 and −2.9 V; the small peaks can be explained with selective adsorption or reorientation of the RTIL ions—as discussed in [12,13]—as well as to some degree of reactivity of the organic [14]; in fact, the RTIL is electrochemically stable to −2.25 V and +1.75 V *vs.* Au QRE: The electrochemical window of stability is therefore ca. 4.0 V. Extending the potential excursion, gives rise to: (i) oxidation processes at +2.0 V, followed by a corresponding reduction event around 1.0 V in the reverse scan and (ii) reduction processes beyond −2.0 V, followed by a corresponding oxidation process around 0 V. The large potential differences between these RTIL breakdown processes and their corresponding reverse reactions indicate the strongly irreversible nature of the corresponding reactions. The CV corresponding to the addition of 1 mM CTDB exhibits an irreversible cathodic peak at ca. −0.5 V in the cathodic-going scan. This cathodic reactivity can be interpreted in terms of denitrilation or diazo bond breaking, on the basis of cognate experiments performed by some of the authors in aqueous solution [10,11] as well as of the literature describing the reaction of similar molecules [15]. Another minor voltammetric feature brought about by the addition of CTDB is the couple of anodic peaks centred at ca. 0.9 and 1.5 V in the anodic-going scan, due to the oxidation of one of the reduction products of CTDB [10]. From our results of [7] we can exclude that this peak is related to the release of $CN^-$ caused by denitrilation. In the presence of $K[Au(CN)_2]$, we found a current density increase from ca. −2.0 V, due to Au electrodeposition (e.g., [16]), followed by a classical mass-transport controlled peak at ca. −2.5 V. The new anodic features that appear after addition of $K[Au(CN)_2]$ can be related to $CN^-$ adsorption, oxidative adsorption and/or formation of Au(I) complexes with RTIL ions, as extensively illustrated in [8]. The CV recorded in the presence of both CTDB and $K[Au(CN)_2]$ is essentially a combination of those obtained with the single reagents, with the difference that the features corresponding to the electrodeposition processes are smaller, as expected from the inhibiting action of the leveller [11]. Mechanistic details on the heterogeneous and homogeneous electrochemical reactions occurring in the different electrolytes are beyond the scope of the present paper.

**Figure 1.** Cyclic voltammetries for polycrystalline Au in contact with [BMP][TFSA]-based electrolytes: CTDB 1 mM, $K[Au(CN)_2]$ 25 mM. Scan rate: 0.1 V·s$^{-1}$, scan started at 0 V *vs.* Au QRE, initial scan direction: cathodic.

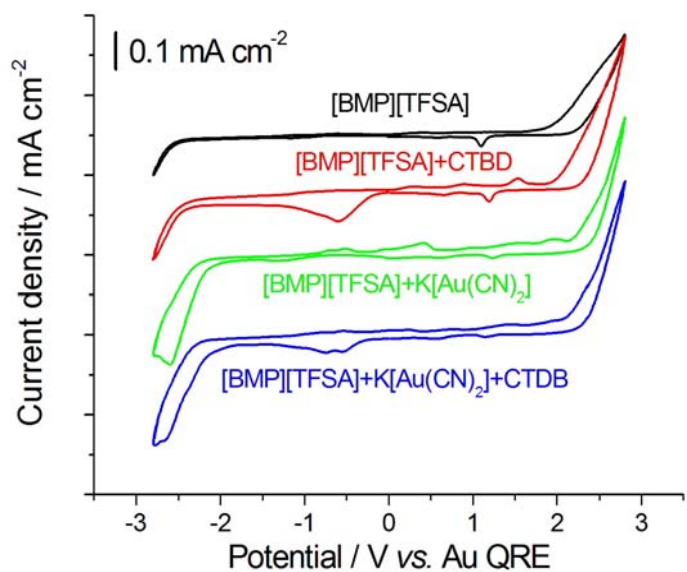

*2.2. In Situ Sum- and Difference-Frequency Generation Spectroscopy (SFG/DFG)*

In order to pinpoint the interfacial behaviour of CTDB at Au electrodes in the absence and presence of Au electrodeposition, we followed—as a function of potential—the CN stretching band, that according to characteristic position, can correspond either to adsorbed nitrile (ca. 2,200 cm$^{-1}$) or to adsorbed CN$^-$ (ca. 2,100–2,150 cm$^{-1}$). The peak position, width, sign and Stark tuning of this adsorbate mode are highly diagnostic of the interfacial structure. Furthermore, since in the case of Au the visible energy lies close to the interband transition, SFG and DFG spectroscopies yield complementary information, sensitive to the orientation of the adsorbed dipole as well as on the non-resonant part of the second-order polarisability $\chi^{(2)}_{NR}$ (for details, see the Appendix), as a result of interference between the resonant and non-resonant contributions. Furthermore, DFG spectroscopy with Au electrodes exhibits a better signal-to-noise ratio owing to the absence of a non-resonant background.

SFG and DFG spectra were recorded in the range −2.5–+2.0 V (Figures 2–4). In the experiments without K[Au(CN)$_2$] the potential was initially set at a value where CTDB is stable and stepped first in the cathodic direction—crossing the critical potential for the cathodic decomposition of CTDB—(Figure 2) and subsequently in the anodic one (Figure 3). The experiments with K[Au(CN)$_2$] were started at a cathodic potential where electrodeposition is active and then shifted in the anodic direction.

**Figure 2.** Potential-dependent SFG spectra of Au in contact with [BMP][TFSA] containing 1 mM CTDB and the corresponding fits: Cathodic-going scan (from +1.50 V to −1.75 V, starting with pristine Au electrode). The potential scan sequence (cathodic-going scan) is indicated by the arrow.

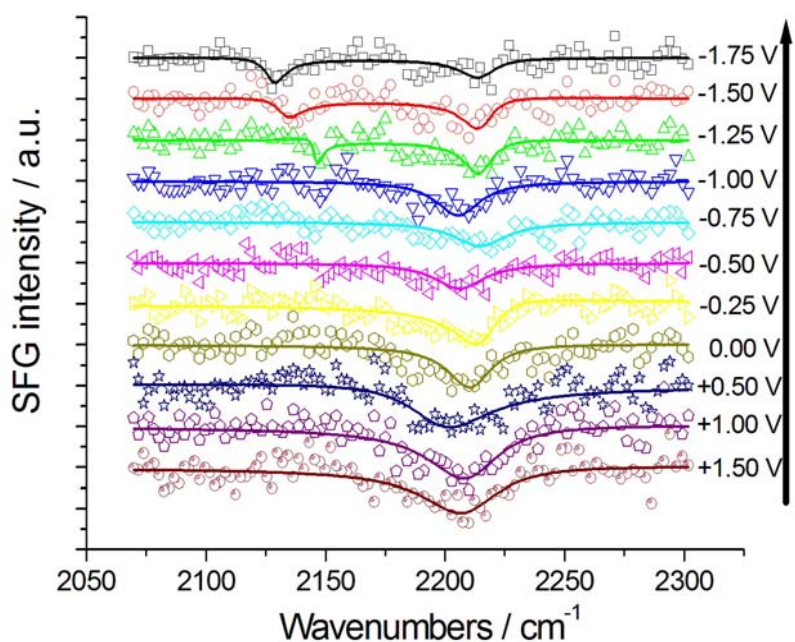

**Figure 3.** Potential-dependent DFG spectra of Au in contact with [BMP][TFSA] containing 1 mM CTDB and the corresponding fits: Anodic-going scan (from −1.75 V to +1.50 V, measurements performed after those reported in Figure 2). The potential scan sequence (anodic-going scan) is indicated by the arrow.

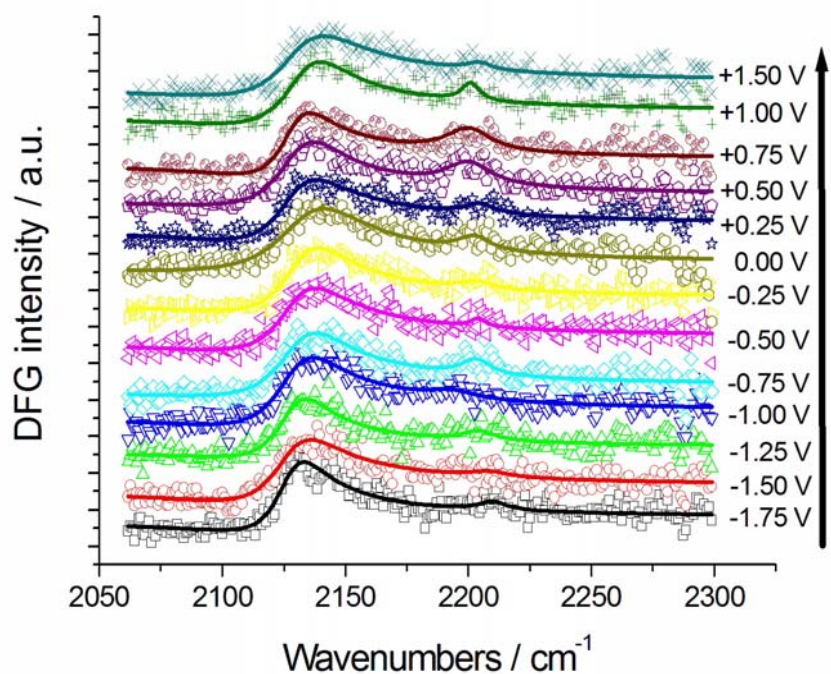

**Figure 4.** Potential-dependent DFG spectra of Au in contact with [BMP][TFSA] containing 1 mM CTDB and 25 mM K[Au(CN)$_2$] and the corresponding fits: anodic-going scan (from −2.50 V to +2.00 V, starting with pristine Au electrode). The potential scan sequence (anodic-going scan) is indicated by the arrow.

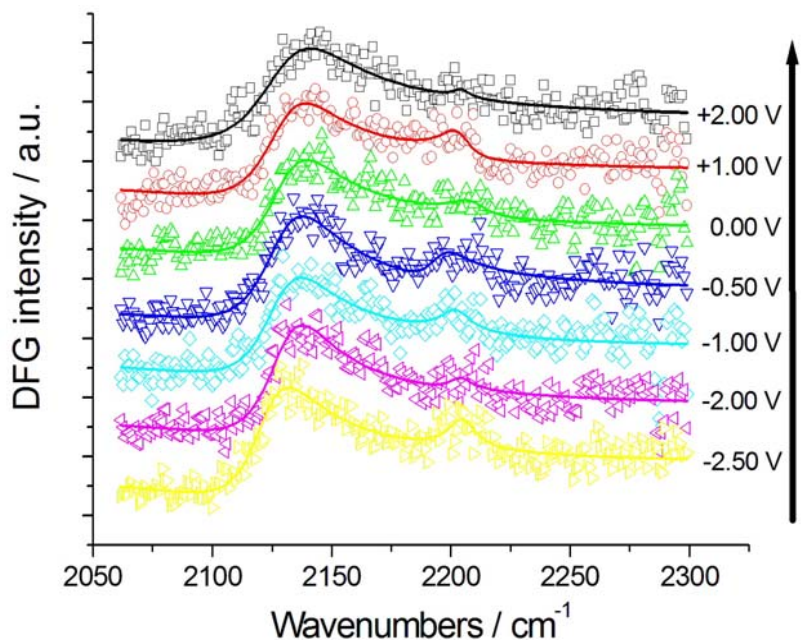

2.2.1. Adsorption and Cathodic Reaction of CTDB in [BMP][TFSA]

Working with a static Au electrode (in the absence of the Au electrodeposition reaction) during the cathodic-going scan (Figure 2), for potentials higher than −1 V, a single, negative SFG C≡N band was detected at 2,210 ± 5 cm$^{-1}$—the peak position does not show a measurable correlation with the applied potential—corresponding to nitrile of CTDB, adsorbed via the C atom [17].

The denitrilation reaction can be related to the shoulder found at ca. −1 V in the cathodic CV peak of the cathodic-going scan (Figure 1). For lower potentials, a negative SFG CN$^-$ band appears in the range 2,130–2,150 cm$^{-1}$, exhibiting a typical Stark tuning, again denoting adsorption via C. As found in the case of aqueous solution, CN$^-$ is the result of cathodic denitrilation of CTDB [11]. It is worth noting that the Stark tuning recorded with incipient formation of adsorbed CN$^-$ released by denitrilation is notably higher than that found with CN$^-$ adsorbed from KCN [7] and K[Au(CN)$_2$] [8] solutions in [BMP][TSFA], as well as that recorded in the anodic-going scan (Figures 3 and 5A) following the cathodic-going one (Figure 2).

**Figure 5.** ν(CN) peak positions ω$_o$ corresponding to CN$^-$ adsorbed at Au in contact with [BMP][TFSA] containing 1 mM CTDB without (**A**, **B**) and with (**C**) 25 mM K[Au(CN)$_2$]. (**A**) Cathodic-going scan, SFG spectra (see Figure 2); (**B**) Anodic-going scan, DFG spectra (see Figure 3). (**C**) Anodic-going scan, DFG spectra (see Figure 4); The vertical grey line indicates the reactivity threshold for CTDB. The error bars correspond to estimated 95% confidence intervals.

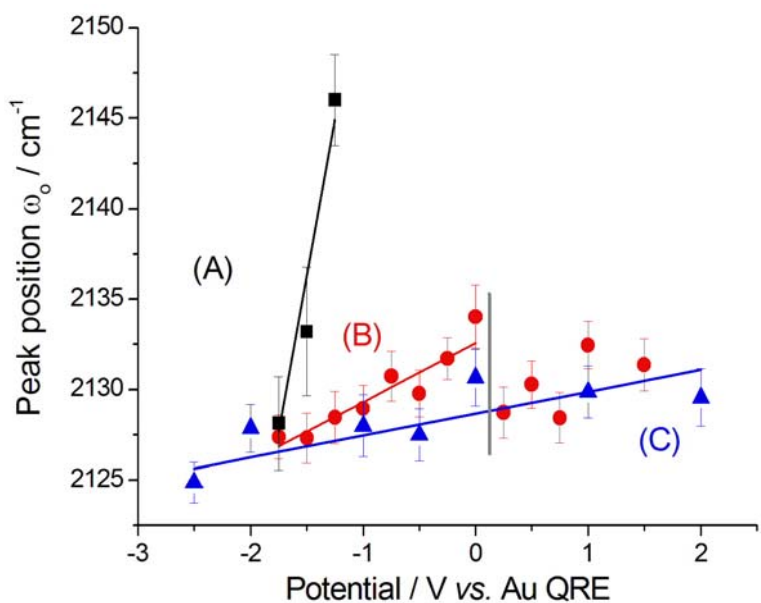

After the cathodic-going scan imposed to an initially pristine electrode, yielding the SFG spectra discussed in the previous paragraph, we switched to the DFG mode and measured an anodic-going scan (Figure 3), with pre-adsorbed CN$^-$ resulting from denitrilation. Pre-adsorbed CN$^-$ is in this case present in the whole investigated potential range and the nitrile band can still be noticed, though with a lower relative intensity with respect to the CN$^-$ one, since the latter species is more strongly adsorbed to Au. The nitrile band position in the return scan was found to be 2,203 ± 4 cm$^{-1}$, essentially the same value as in the forwards one. Figure 5B shows the peak positions of the adsorbed CN$^-$ band, exhibiting a Stark tuning for potentials more cathodic than the CTDB reactivity threshold (see Figure 1) and an approximately constant value (2,131 ± 2 cm$^{-1}$) for higher potentials. The estimated Stark tuning value (3.3 ± 0.5 cm$^{-1}$ V$^{-1}$) is essentially identical to that found in KCN-containing solutions [7] (SFG 2.9 ± 0.6 cm$^{-1}$ V$^{-1}$, DFG 3.6 ± 0.2 cm$^{-1}$ V$^{-1}$).

The nitrile bandwidth $\Gamma$ exhibits a clear decreasing trend in the cathodic-going scan (Figure 6A), while it is essentially constant in the anodic-going one (Figure 6B), denoting a selection of the type of adsorbed nitrile, probably corresponding to the moiety belonging to one of the cleavage products pinpointed in [10]. The CN$^-$ bandwidth $\Gamma$ is smaller in the initial stages of pseudo-halide formation in the cathodic-going scan (Figure 7A); once it is formed, the bandwidth follows a trend (Figure 7B) that is very similar to that of nitrile, though with slightly higher absolute values. The explanation given above for the increase of $\Gamma$ with potential in the case of nitrile, can be conjectured to hold also for CN$^-$.

**Figure 6.** Peak width $\Gamma$ corresponding to nitrile adsorbed at Au in contact with [BMP][TFSA] containing 1 mM CTDB. (**A**) Cathodic-going scan, SFG spectra (see Figure 2); (**B**) Anodic-going scan, DFG spectra (see Figure 3). The error bars correspond to estimated 95% confidence intervals.

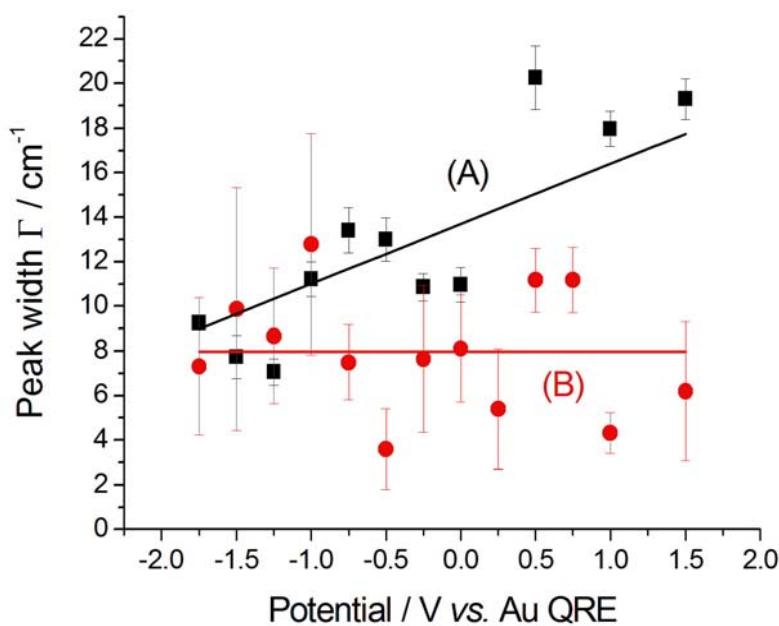

The height $\beta$ of the nitrile peak shows a general anticorrelation with the applied potential, due to reactivity at cathodic polarisations (Figure 8). The decrease in $\beta$ values found at at −1.75 V between the cathodic- and anodic-going scans is due to the fact that we kept the potential applied while switching from SFG to DFG and the reaction, leading to CTDB consumption, continued. Competition for adsorption with CN$^-$ on Au corroding at high anodic potentials gives rise to a drop of $\beta$. Since CN$^-$ forms during the cathodic-going scan, the peak height $\beta$ of CN$^-$ appears only at sufficiently high cathodic polarisations (Figure 9A), then it grows during the anodic-going scan (Figure 9B) according to the customary potential-dependent adsorption behaviour of this pseudo-halide on Au [17]. Again, as in the case of nitrile (Figure 8), the discontinuity in the value of $\beta$ at −1.75 V between the cathodic- and anodic-going scans is due to the build-up of the reaction product during the holding period corresponding to switching from SFG to DGF.

**Figure 7.** Peak width Γ corresponding to CN⁻ adsorbed at Au in contact with [BMP][TFSA] containing 1 mM CTDB without (**A**, **B**) and with (**C**) 25 mM K[Au(CN)$_2$]. (**A**) Cathodic-going scan, SFG spectra (see Figure 2); (**B**) Anodic-going scan, DFG spectra (see Figure 3); (**C**) Anodic-going scan, DFG spectra (see Figure 4). The error bars correspond to estimated 95% confidence intervals.

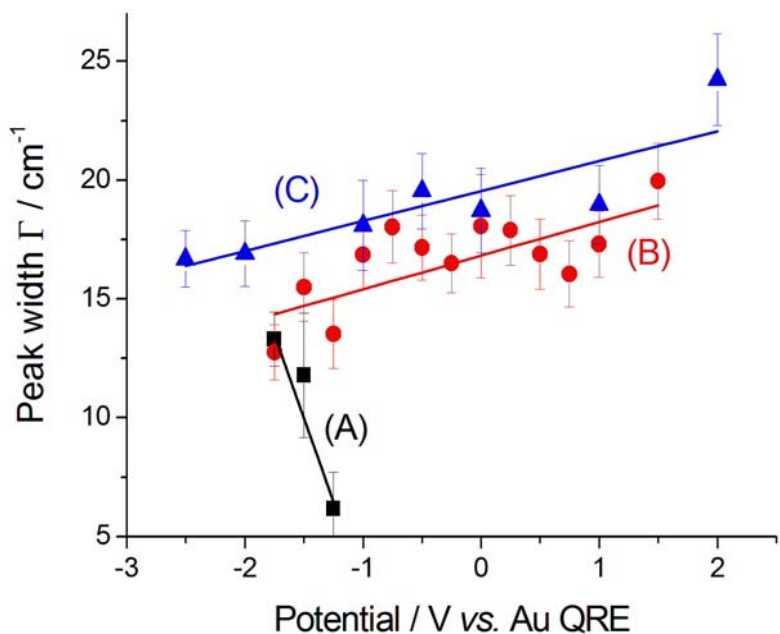

**Figure 8.** Peak height β corresponding to nitrile adsorbed at Au in contact with [BMP][TFSA] containing 1 mM CTDB. (**A**) Cathodic-going scan, SFG spectra (see Figure 2); (**B**) Anodic-going scan, DFG spectra (see Figure 3). The error bars correspond to estimated 95% confidence intervals.

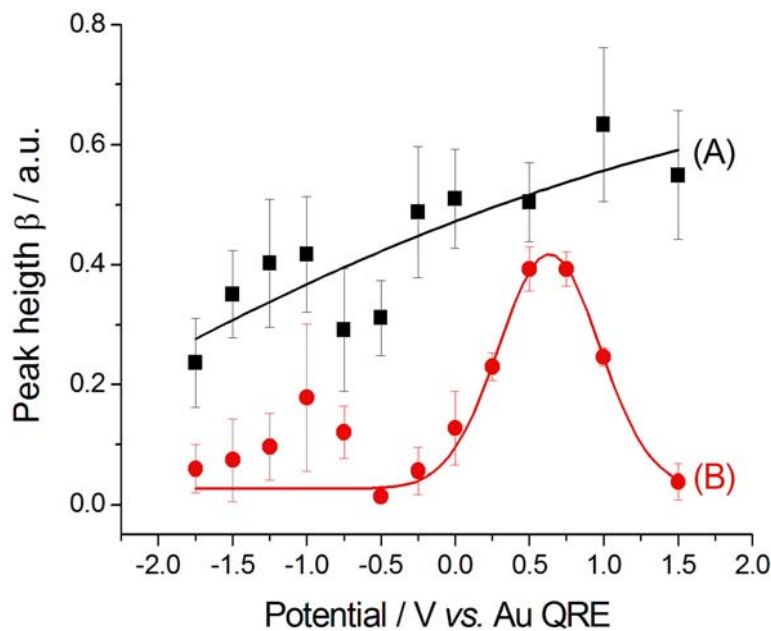

**Figure 9.** Peak height β corresponding to CN⁻ adsorbed at Au in contact with [BMP][TFSA] containing 1 mM CTDB without (**A**, **B**) and with (**C**) 25 mM K[Au(CN)$_2$]. (**A**) Cathodic-going scan, SFG spectra (see Figure 2); (**B**) Anodic-going scan, DFG spectra (see Figure 3); (**C**) Anodic-going scan, DFG spectra (see Figure 4). The error bars correspond to estimated 95% confidence intervals.

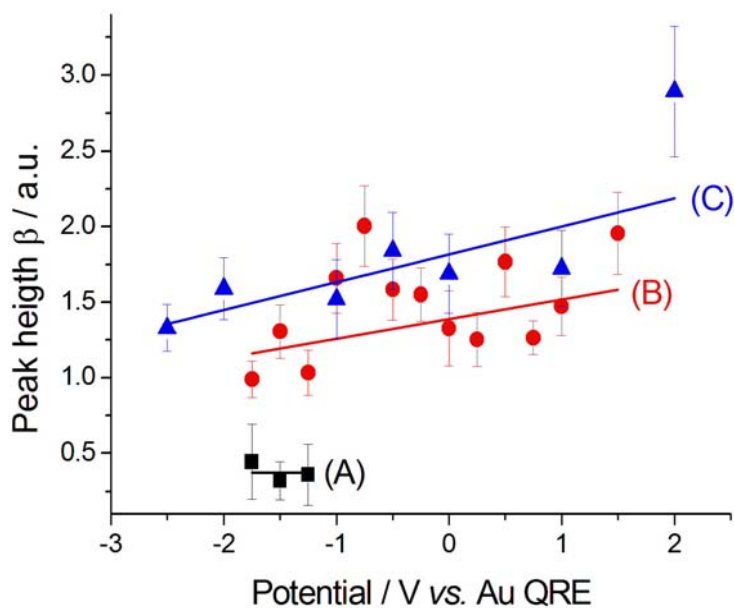

The value of the interference parameter δ (for details, see the Appendix) does not exhibit a definite trend in either scan and it fluctuates close to zero: −0.018 ± 0.025. The potential-dependence of the non-resonant parameter α estimates from SFG measurements is shown in Figure 10.

**Figure 10.** Non-resonant parameter α for Au in contact with [BMP][TFSA] containing 1 mM CTDB without (A, B) and with (C) 25 mM K[Au(CN)$_2$]. (**A**) Cathodic-going scan, SFG spectra (see Figure 2); (**B**) Anodic-going scan, DFG spectra (see Figure 3); (**C**) Anodic-going scan, DFG spectra (see Figure 4). The error bars correspond to estimated 95% confidence intervals.

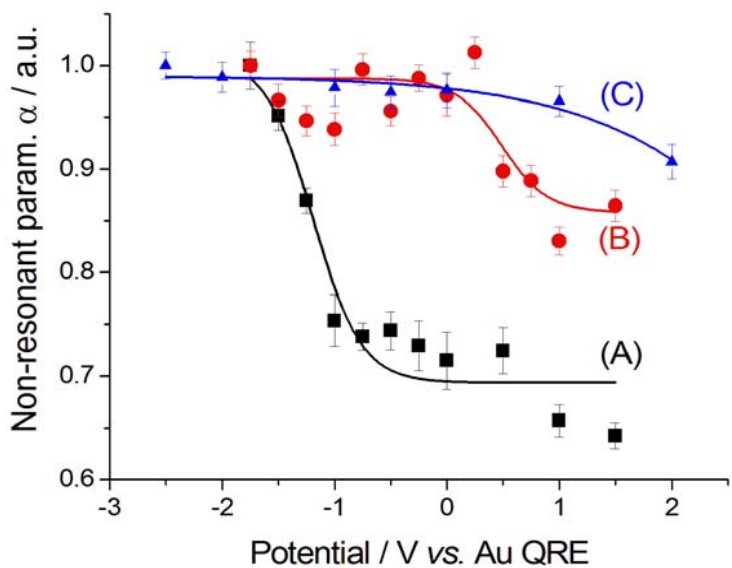

We report data normalised on the value of the parameter estimated from the spectra recorded at the most cathodic values, because the absolute value of α is affected by the specific configuration of the optical setup, as detailed in [18]. A clear potential-dependent, hysteretic behaviour can be observed, with higher values at cathodic potentials. Even though a mechanistic justification of this behaviour is beyond the scope of this paper, some correlation can be noticed between the discontinuity in plot (A) and the potential of formation of adsorbed $CN^-$ in the cathodic-going scan (ca. −1,25 V); the hysteresis, showing higher α values in the anodic-going scan (B) might be related to the presence of adsorbed $CN^-$, the drop at higher potentials might correlate with the CV features discussed in Section 2.1.

2.2.2. Electrodeposition of Au from a [BMP][TFSA]-based solution containing CTDB and $K[Au(CN)_2]$

The DFG spectra recorded during an anodic-going scan starting at −2.5 V are reported in Figure 4. The spectral patterns closely resemble those measured with the solution not containing $Au(CN)_2^-$, after release of $CN^-$ in the cathodic-going scan (Figure 4) and exhibits two ν(CN) peaks corresponding to $CN^-$ and nitrile in approximately the same positions.

As far as the nitrile peak is concerned, its parameters are essentially independent on the potential: (i) the peak position $\omega_o$ is 2,204 ± 3 $cm^{-1}$, the same value found in the absence of electrodeposition (Section 2.2.1); (ii) the peak width Γ is 7.6 ± 1.9 $cm^{-1}$, very close to that estimated in the anodic-going scan with just CTDB in the solution; (iii) the peak height β is 0.10 ± 0.06, very close to the values recorded in the absence of $Au(CN)_2^-$, but it does not seem to exhibit the maximum shown is Figure 8B.

The $CN^-$ peak exhibits a very small Stark tuning of 1.2 ± 0.6 $cm^{-1}$ $V^{-1}$ (Figure 5C). This value is smaller than that found without $K[Au(CN)_2]$ in the same solvent and notably smaller than that measured for a [BMP][TFSA] solution with $K[Au(CN)_2]$, but without CTDB: 11.2 ± 1.1 $cm^{-1}$ $V^{-1}$ [8]. This notable reduction in the Stark tuning in the case of coadsorption of $CN^-$ and CTDB-related species reasonably correlates with the levelling activity of the organic. The behaviour of the peak width Γ (Figure 7C) and height β (Figure 9C) is very similar to that found in the absence of electrodeposition and the same comments of Section 2.2.1 apply also here. Also the potential-dependence of the non-resonant parameter α behaves as in the anodic-going scan measured from the electrolyte without the Au(I) cyanocomplex (Figure 10C). These results show that the interfacial behaviour of $CN^-$ at the Au surface, both is the presence and in the absence of Au electrodeposition—as studied by DFG—is essentially the same: since in previous investigations we have found notable effects of the electrodeposition process on the *in situ* spectroscopic behaviour of $CN^-$, due to the optical properties of nano-sized crystallites [16,19,20] or layered structures [21] forming by electrodeposition, the present result can be correlated to the excellent levelling properties of CTDB, giving rise to growth of Au with spectroscopic properties that are not distinguishable from those of polished polycrystalline Au.

## 3. Experimental

The basic Au electrodeposition bath was the same as described in [8]: Details on solution preparation and handling are also provided in [8]. The composition was 0.025 M $K[Au(CN)_2]$ (Engelhard) solution in [BMP][TFSA] (99% Iolitec). To this bath we added CTDB 1 mM (Maybridge). The working electrode was polycrystalline Au disk of 8 mm diameter and 3 mm thickness, treated by flame annealing, as detailed in [7]. The quasi-reference (QRE) and counter electrodes were Au wire, as customary in the literature [1]: All potentials are reported *vs.* Au QRE. The thin-layer cell is described in detail in [22]. The optical setup using the infrared optical parametric oscillator (IR-OPO) is detailed in [22]: Briefly, p-polarised tunable IR is delivered between 2.7 and 6 μm with energy resolution of 2 cm$^{-1}$. The p-polarised VIS is a doubled Nd:YAG. Spectral modelling and data-processing methods are illustrated below.

*3.1. Single-Resonance Model for SFG/DFG*

According to the classical approach of [17], SFG/DFG spectra exhibiting a single peak can be modelled as:

$$I \propto \left|\chi^{(2)}\right|^2 \tag{1}$$

where:

$$\chi^{(2)} = \chi^{(2)}_{NR} + \chi^{(2)}_R = (a + ib) + \frac{A}{(\omega - \omega_o) \pm i\Gamma} \tag{2}$$

and the subscripts *R* and *NR* stand for "resonant" and "non-resonant" respectively, *a* and *b* are the free- and bound-electron contributions, *A* is the resonator strength, $\omega_o$ is the corresponding resonant frequency and *Γ* its width. Elaborating on Equation (2) and defining: $\omega - \omega_o = x$, by simple algebra it can be shown that:

$$\left|\chi^{(2)}\right|^2 = \alpha + \frac{\beta}{x^2 + \Gamma^2} \cdot (1 + \delta \cdot x) \tag{3}$$

where: $\alpha = a^2 + b^2$, $\beta = A \cdot (A \mp 2b\Gamma)$ and $\delta = \frac{2a}{A \mp 2b\Gamma}$. The form of Equation (3) - with $\omega_o$, $\alpha$, $\beta$ and $\delta$ as fit parameters, ensures minimal parameter correlation for the NLLS fitting procedure [9]. Furthermore, Equation (3) allows a straightforward analytical interpretation of the parameter set $\{\omega_o, \Gamma, \alpha, \beta, \delta\}$: (i) $\alpha$ is the background far from the resonance; (ii) $\omega_o$ is the peak position in case of a pure Lorentzian lineshape; (iii) $\frac{1}{x^2 + \Gamma^2}$ is a Lorentzian lineshape; (iv) $\frac{\beta}{\Gamma^2}$ is the height of the Lorentzian peak above background; (iv) *Γ* is the peak width at half maximum; (v) $(1 + \delta \cdot x)$ can be understood as a linear approximation of a function describing the distortion of the lineshape from a pure Lorentzian (of course, a pure Lorentzian is obtained if $\delta = 0$).

*3.2. N-Resonance Model for SFG/DFG*

Elaborating on Equation (3), it can be straightforwardly proved that:

$$\chi^{(2)} = a + ib + \sum_{j=1}^{N} \frac{A}{(\omega - \omega_{o,j}) \pm i\Gamma_j} \tag{4}$$

defining: $\omega - \omega_{o,j} = x_j$, by lengthy, but otherwise simple algebra one can derive:

$$|\chi^{(2)}|^2 = \alpha + \sum_{j=1}^{N} \frac{\beta_j}{x_j^2 + \Gamma_j^2} \cdot (1 + \delta_j \cdot x_j) + \frac{1}{4a^2} \cdot \sum_{j,k=1}^{N} \varepsilon_{jk} \cdot \frac{x_j \cdot x_k + \Gamma_j \cdot \Gamma_k}{(x_j^2 + \Gamma_j^2) \cdot (x_k^2 + \Gamma_k^2)} \quad (5)$$

where: $\alpha = a^2 + b^2$, $\beta_j = A_j \cdot (A_j \mp 2b\Gamma_j)$, $\delta_j = \frac{2a}{A_j \mp 2b\Gamma_j}$ and $\varepsilon_{jk} = \delta_j \cdot \delta_k$. The additional terms $\varepsilon_{jk}$ express the pairwise interference between couples of resonances.

*3.3. Identification of a Guess Set of Parameters by Graphical Approach and Linear Least-Squares*

In this section, the discussion is limited to the case N = 1 for DFG, but the same approach can be followed for a general N and both SFG and DFG, provided the resonances are separated (*i.e.*, $x_j - x_k \gg \Gamma_j \approx \Gamma_k$; in fact, it can be proved that if $x_j - x_k > 3\Gamma$ and $\Gamma_j \approx \Gamma_k \approx \Gamma$, $\left( |\chi^{(2)}|^2_{indep} - |\chi^{(2)}|^2_{coupled} \right) / |\chi^{(2)}|^2_{indep} < 0.1$, where the subscripts "indep" and "coupled" refer to independent *vs.* couples resonances). The authors were not able to find a general approach to the problem of guess set identification in the case of strongly interacting resonances.

With these provisos, following the analytical interpretation of Equation (3) discussed in Section A1 and by inspection of its form, it is straightforward to assign graphically a set of values for the subset $\{\omega_o, \Gamma, \alpha, \beta\}$ of the parameter set $\{\omega_o, \Gamma, \alpha, \beta, \delta\}$. Once the subset $\{\omega_o, \Gamma, \alpha, \beta\}$ is assigned, parameter $\delta$ can be obtained by solving the following linear least-squares problem: (i) the experimental (measured) array $\{|\chi^{(2)}|^2\}_{meas} \overset{\Delta}{=} y_{meas}$ is transformed into $\tilde{y}_{meas} = (y_{meas} - \alpha) \cdot \frac{x^2 + \Gamma^2}{\beta}$, (ii) we adopt the transformed model: $\tilde{y}_{comp} = 1 + \delta \cdot x$ and (iii) identify $\delta$ with a linear least-squares fit of data $\tilde{y}_{meas}$ with model $\tilde{y}_{comp}$. In this way, we have generated a guess set of parameters $\{\omega_o, \Gamma, \alpha, \beta, \delta\}_{guess}$ that is expected to be reasonably close to the global minimum (or, alternatively, to an NGM [23] whose physical meaning is clear). At this point, the original data set $y_{meas}$ can be fitted with full model of Equation (2) by NLLS, employing $\{\omega_o, \Gamma, \alpha, \beta, \delta\}_{guess}$ as the starting parameter choice. The NLLS code hopefully will seek a sound minimum (in the sense defined above) along a reasonably hyperparabolic objective function.

*3.4. Recovery of the "Physical" Parameter Set from the "Minimal-Correlation" Parameter Set*

In this section we discuss the special case of N = 1 for DFG. A similar approach can be taken in the general case, again, provided the resonances are sufficiently separated, in the sense discussed in Section A3. The original parameter set $\{A, a, b\}$ can be recovered from the transformed parameter set $\{\alpha, \beta, \delta\}$ through the algebraic manipulations explained below.

(i) Since, in the case of Au, $a < b$ for Au, we can take: $\alpha = a^2 + b^2 \cong b^2$, whence: $b \cong \sqrt{\alpha}$.

(ii) It is possible to use the approximation (i) to estimate *A* from β as follows:

$\beta \overset{\Delta}{=} A \cdot (2b\Gamma + A) \cong A \cdot (2\sqrt{\alpha}\Gamma + A)$, whence: $A_{1,2} = -\sqrt{\alpha} \cdot \Gamma \pm \sqrt{\alpha \cdot \Gamma^2 + \beta}$. Since physical solutions ought to be positive, it follows that: $A = \sqrt{\alpha \cdot \Gamma^2 + \beta} - \sqrt{\alpha} \cdot \Gamma$.

(iii) At this point, it is possible to estimate *a* from $\delta$:

$$\delta \stackrel{\Delta}{=} \frac{2a}{2b\Gamma + A} \cong \frac{2a}{2\sqrt{\alpha \cdot \Gamma} + A} = \frac{2a}{\sqrt{\alpha\Gamma^2 + \beta} + \sqrt{\alpha\Gamma^2}}, \text{ whence: } a \cong \frac{\delta}{2} \cdot \left( \sqrt{\alpha\Gamma^2 + \beta} + \sqrt{\alpha\Gamma^2} \right)$$

(iv) Once an estimate of *a* is available, it is possible to produce a better estimate of *b* by using the exact expression: $b \cong \sqrt{\alpha - a^2}$. This updated value of *b* can, of course be used at point (ii) above and following to set up an iterative scheme.

It is worth noting in conclusion that, in any case, the "minimal-correlation" parameter set can be used directly for a meaningful physical discussion of spectral results and it is not always necessary to go back to the original parameter set of Equation (2).

## 4. Conclusions

In this paper we report on the electrochemical adsorption of CTDB from an [BMP][TFSA] RTIL solution on Au. The Au electrode is either metallographically polished polycrystalline Au or the dynamic surface resulting from ongoing growth by electrodeposition. This study is based on electrochemical measurements and *in situ* SFG/DFG non-linear spectroscopy. As a function of applied potential and electrode history, the adsorbed species are either a nitrile moiety of CDTB or coadsorbed nitrile and CN⁻. In the solution containing only CTDB, CN⁻ derives from cathodic decomposition of the organic. In the electrodeposition bath, CN⁻ is also released as a result of the reductive decomposition of the Au(I) cyanocomplex. Quantitative analyses of potential-dependent SFG/DFG spectra have disclosed details on the adsorption modes of CTDB and CN⁻ at the Au surface and on their mutual interaction. On the basis of Stark tuning measurements, CTDB notably lowers the interaction of CN⁻ with the growing Au surface: This behaviour correlates with a beneficial effect on electrodeposit quality, in terms of morphological control. Apart from the potential-dependent ν(CN⁻) resonance position, in the presence of CTDB the electrodeposition process was found to have a limited bearing on the vibrational properties of the coadsorbates as well as on the electronic properties of the metal substrate, proving that the empirically observed levelling effect of CTDB in electrodeposition has a molecular correlate in the fact that the optical properties of the Au surface are the same for a polished sample and for the material growing by electrochemical reduction.

## Acknowledgements

Continuous, high-standard technical assistance with electrochemical measurements and preparation of spectroelectrochemical cells is gratefully acknowledged to Francesco Bogani, Department of Innovation Engineering, University of Salento.

*Sample Availability*: Samples of the compounds quoted in the text are available from the mentioned vendors.